\documentclass[aps,superscriptaddress,showpacs,prd,nofootinbib,eqsecnum]{revtex4-1}

\usepackage{xcolor}
\usepackage{epsfig}
\usepackage{caption}
\usepackage{subcaption}
\usepackage{graphicx}  
\usepackage{dcolumn}   
\usepackage{bm}        
\usepackage{amssymb}   

\usepackage[colorlinks=true,citecolor=blue,linkcolor=blue,urlcolor=blue]{hyperref}

\def\nn{\nonumber}
\def\beq{\begin{equation}}
\def\eeq{\end{equation}}
\def\be{\begin{equation}}
\def\ee{\end{equation}}
\def\bea{\begin{eqnarray}}
\def\eea{\end{eqnarray}}
\def\nnb{\nonumber}

\def\ms{\overline{\rm MS}}

\def\t{\tilde }

\def\eps{\epsilon}

\def\ds{\displaystyle}
\def\sumint{\hbox{$\sum$}\!\!\!\!\!\!\!\int}
 
\newcommand{\lsim}{\raisebox{-0.13cm}{~\shortstack{$<$ \\[-0.07cm] $\sim$}}~}
\newcommand{\gsim}{\raisebox{-0.13cm}{~\shortstack{$>$ \\[-0.07cm] $\sim$}}~}

\begin{document}

\title{Renormalization group optimized $\lambda \phi^4$ pressure at next-to-next-to-leading order}

\author{Lo\"{\i}c Fernandez}

\affiliation{Laboratoire Charles Coulomb (L2C), UMR 5221 CNRS-Univ. Montpellier, Montpellier, France}
\author{Jean-Lo\"{\i}c Kneur}

\affiliation{Laboratoire Charles Coulomb (L2C), UMR 5221 CNRS-Univ. Montpellier, Montpellier, France}

\begin{abstract}
We investigate the renormalization group optimized perturbation theory (RGOPT) at the next-to-next-to-leading order (NNLO) 
for the thermal scalar field theory. From comparing three thus available successive RGOPT orders we illustrate 
the efficient resummation and very good apparent convergence properties of the method. In particular
the remnant renormalization scale dependence of thermodynamical quantities
is drastically improved as compared to both
standard perturbative expansions and other related 
resummation methods, such as the screened perturbation theory. 
Our present results thus constitute a useful first NNLO illustration in view of
NNLO applications of this approach to the more involved thermal QCD.
\end{abstract}

\maketitle
\section{Introduction}
For thermodynamical quantities at equilibrium, and a weakly coupled theory, one could hope at first that a perturbative 
expansion would give reliable results. In contrast, as is well-known, for a massless theory infrared divergences 
spoil a naive perturbation theory (PT) approach in thermal field theories. Even though those infrared divergences 
can be efficiently resummed (see {\it e.g.} \cite{Trev,kapusta-gale,la-vu} for reviews), 
it leads to nonanalytical terms in the coupling, that happen to 
give poorly convergent successive orders, even when pushed to the 
highest perturbative order available, furthermore with increasingly sizeable remnant scale dependence. This situation is notoriously illustrated for asymptotically free thermal QCD, 
where however lattice simulations (LS) offer a powerful genuinely nonperturbative alternative bypassing those issues. 
At least so far LS has been very successful in the description of the nonperturbative physics of the QCD phase transitions 
at finite temperatures and near vanishing or small baryonic densities~\cite{QCDlattT}.   
Nevertheless, the famous numerical sign problem \cite{signpb} at finite density (equivalently at finite chemical potential),  
prevents LS to successfully describe compressed baryonic matter at sufficiently high densities and to explore a large part
of the QCD phase diagram. 
More generally despite the very success of LS it is still highly desirable to explore more analytical  
improvements of thermal PT. Accordingly many efforts have been devoted in the past to overcome
the generically observed issues of poor PT convergence.
Apart from the most important case of QCD, the above mentioned behavior is generic for any thermal quantum field theory, 
and typically the scalar $\lambda \phi^4$ interaction is often used as a simpler model to study new alternative approaches.
Although the $\lambda \phi^4$ model is not asymptotically free, it shares some important features with thermal QCD. 
Typically the dynamical generation of a thermal screening mass $m_D\sim \sqrt{\lambda} T$ impacts 
the relevant expansion of thermodynamical quantities such as the pressure, that exhibit 
weak expansion terms $\lambda^{(2p+1)/2}, p\ge 1$.

Various approximations attempting to more efficiently resum thermal perturbative expansions 
have been developed and refined over the years, typically the screened perturbation theory (SPT)\cite{SPTearly,SPT},
the nonperturbative renormalization group (NPRG)\cite{NPRG} approach, the two-particle irreducible (2PI) 
formalism\cite{2PI,2PI3loop,2PI4loop}, 
or other approaches\cite{Blaizot-Wschebor}. 
In particular for the $\lambda \phi^4$ model, SPT essentially 
redefines the weak expansion about a quasiparticle mass, avoiding in this way infrared divergences from the start. 
It has been investigated up to three-loop\cite{SPT3l,SPT3lnext,OPT3l} and even four-loop orders\cite{spt4L}. 
SPT may also be viewed as a particular case, in the thermal context, of the so-called optimized perturbation theory 
(OPT)\footnote{OPT and its many variants appear under different names in the literature~\cite{OPT-LDE,odm,pms}.}, 
in which more generally the (thermal or nonthermal) perturbative expansion is redefined about an unphysical 
test mass parameter, fixed by a variational prescription, that provides a resummation of perturbative expansion. 
The generalization of SPT tailored to treat the much more involved thermal gauge theories\cite{HTL}, 
the Hard Thermal Loop perturbation theory (HTLpt)\cite{htlpt1},  
has been pushed to three-loop order~\cite{HTLPT3loop,HTLPTMU}. 
The three-loop results\cite{HTLPTMU} agreement with LS is quite remarkable 
down to about twice the critical temperature, for a renormalization scale choice $\sim 2\pi T$. However, both SPT and HTLpt
exhibit a very sizeable remnant scale dependence at NNLO, that definitely call for further improvement. \\
More recently, OPT at vanishing temperatures and densities was extended to the so-called 
renormalization group optimized perturbation theory (RGOPT)\cite{rgopt1,rgopt_alphas}. 
The basic novelty is that it restores perturbative RG invariance
at all stages of calculations, in particular when fixing
the variational mass parameter, by solving
the (mass) optimization prescription consistently with
the RG equation. At vanishing temperatures and densities it has given precise first principle 
determinations~\cite{rgopt_alphas}
of the basic QCD scale ($\Lambda_{\overline {\rm MS}}$) or related coupling $\alpha_S$, and of the
quark condensate~\cite{rgopt_qq,rgopt_qq5}. 
The RGOPT was extended at finite temperatures for the scalar $\lambda \phi^4$ in \cite{prl_phi4,prd_phi4} 
and for the non-linear sigma model (NLSM) in \cite{nlsm},
showing how it substantially 
reduces the generic scale dependence and convergence problem of thermal perturbation theories at 
increasing perturbative orders. More recently the RGOPT in the quark sector contribution 
to the QCD pressure was investigated at NLO, for finite densities and vanishing temperatures\cite{prdCOLD},
and at finite temperature and density\cite{rgopt_NLOQCD1,rgopt_NLOQCD2}, leading to drastic improvements
with respect to both perturbative QCD and HTLpt, specially at nonzero temperature. 
In the present work, as a first step to investigate the RGOPT in thermal theories beyond NLO, 
we explore the three-loop order (NNLO) for the technically
simpler scalar $\lambda \phi^4$ model pressure.
We investigate the stability and convergence properties of the method,
also assessing the remnant renormalization scale dependence improvement as compared to standard PT, and to SPT. 

The paper is organized as follows. In Sec. II we briefly review the standard thermal perturbative pressure of 
the $\lambda \phi^4$ model
up to NNLO, to set our conventions and basic expressions that serve as a starting point for our construction.
In Sec. III we recall the main ingredients of the RGOPT construction and some previous NLO results from \cite{prl_phi4,prd_phi4}.
Our main new NNLO results are derived and illustrated in Sec IV, while finally some conclusions and outlook are given in Sec. V.
Some additional technical ingredients can be found in three appendixes.
\section{Review of standard (massive) thermal perturbative expansion} 
We consider the Lagrangian for one neutral scalar field with a quartic interaction,
\be
{\cal L} = \frac{1}{2} \partial_\mu \phi \partial^\mu \phi -\frac{m^2}{2}\phi^2 -\frac{\lambda}{4!} \phi^4\,\,,
\label{Lphi4}
\ee
where a generic mass term $m$ is arbitrary at this stage. 
%
\begin{figure}[h!]
\epsfig{figure=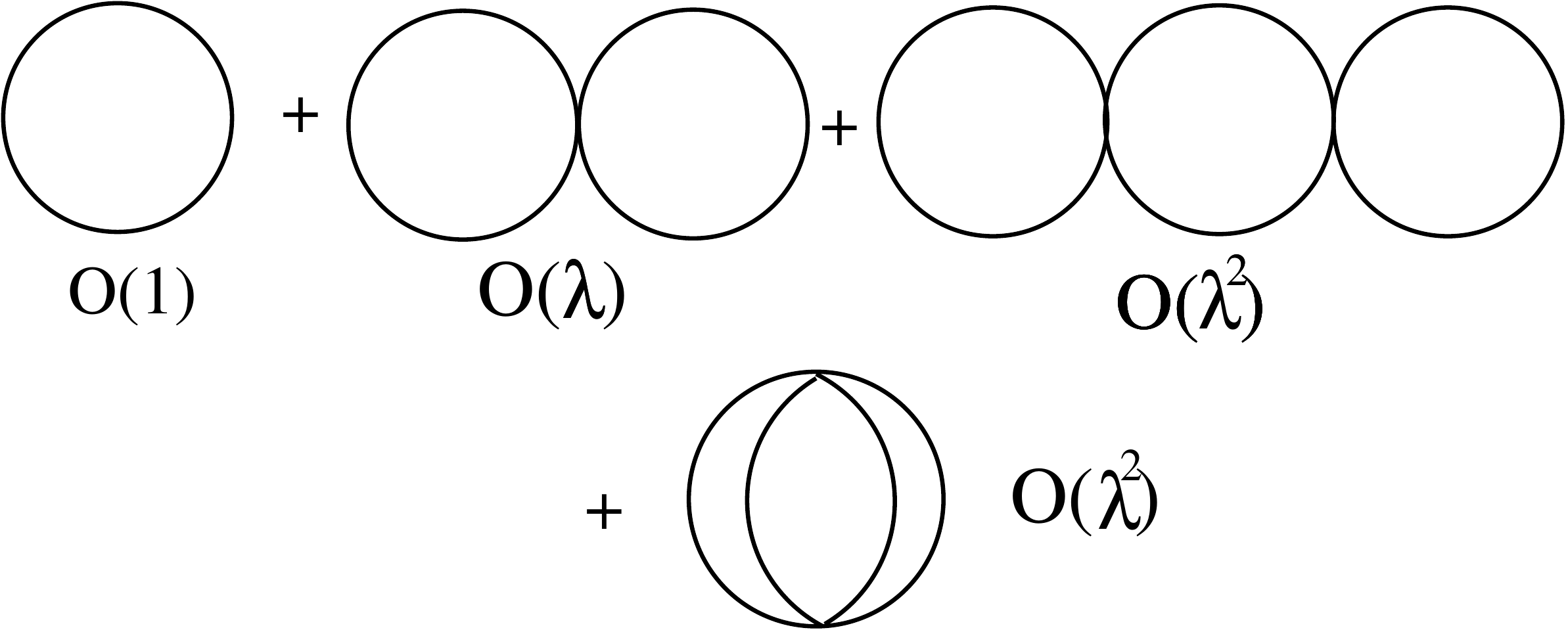,width=8cm,height=3.cm}
\caption{ Free energy diagrams up to NNLO in $\lambda \phi^4$ model. }
\label{Fgraph}
\end{figure}
\subsection{Free energy up to NNLO}
We first recall the result\cite{kapusta-gale,SPT3l} 
for the two-loop free energy (equivalently minus the pressure) including a mass term,
corresponding to the first two graphs~\footnote{In Fig.\ref{Fgraph} counterterm graphs are omitted for simplicity.}
in Fig.\ref{Fgraph},
\be
{\cal F}_0 = \frac{1}{2}
\sumint_P \ln ({\bf P}^2 +m_0^2)  +
\frac{\lambda_0}{8} \left(\sumint_P \frac{1}{{\bf P}^2+m_0^2} \right)^2 +{\cal F}_0^{2l,\rm ct},
\label{2lbasic}
\ee
where in the imaginary time formalism ${\bf P}^2 = \omega_n^2 + {\bf p}^2 $, $\omega_n=2\pi T n$ $(n=0,1,\cdots)$ 
represents the bosonic Matsubara frequencies. 
The sum-integral in Eq.(\ref{2lbasic}) is defined as usual as the sum over Matsubara frequencies times remaining 
integration over three-momentum, 
using dimensional regularization and the $\ms$ renormalization scheme:
\be
\sumint_P \equiv T \sum_n (\frac{\mu^2 e^{\gamma_E}}{4\pi})^\epsilon \,\int \frac{d^{3-2\epsilon} {\bf p}}{(2\pi)^{3-2\epsilon}} .
\label{dimreg}
\ee
The three-loop contributions involves the basic last two graphs in Fig. \ref{Fgraph} and read 
\bea
&& {\cal F}^{3l}_0 = -\frac{\lambda_0^2}{48}
\left [  3\left(\sumint_p \frac{1}{{\bf p}^2+m_0^2} \right)^2 \sumint_q \frac{1}{({\bf q}^2+m_0^2)^2} \right . \nnb \\
&& \hspace{-.4cm} \left. + \sumint_{p q r} \frac{1}{({\bf p}^2+m_0^2)({\bf q}^2+m_0^2)({\bf r}^2+m_0^2)({\bf p+q+r}^2+m_0^2)}
\right ]  \nnb \\
&& +{\cal F}_0^{3l,\rm ct} .
\label{3lbasic}
\eea
We recall that the renormalization is most easily performed as follows. 
First, applying mutiplicative renormalization to the (bare) coupling and mass in expressions above, as
\be
\lambda_0 =  \lambda Z_\lambda, \;\; m_0=m\,Z_m ,
\ee
where $Z_\lambda, Z_m$ are the standard coupling and mass 
counterterms for the massive $\lambda \phi^4$ model, given for completeness
respectively in Eqs.(\ref{Zlam2l}),(\ref{Zm2l}) in Appendix \ref{appRG}. Those are expanded for the 
relevant three-loop order case here to perturbative order $\lambda^2$.
The remaining divergences are then removed by an additive renormalization from the 
vacuum energy counterterms\cite{SPT3l},
formally represented as ${\cal F}_0^{2l,\rm ct}$, ${\cal F}_0^{3l,\rm ct}$ in the above expressions, 
and also given for completeness in Eq.(\ref{Zvac}) in Appendix \ref{appRG}.

Once the mass, coupling, and vacuum energy
counterterms have been accounted  
to cancel the original divergences,
one obtains  the ($\ms$-scheme) renormalized free energy. The one- and two-loop contributions read~\cite{SPT3l,kapusta-gale}:
\be
 (4\pi)^2 {\cal F}^{NLO} = {\cal E}_0 -\frac{1}{8} m^4 \left( 3+2 L \right) 
-\frac{T^4}{2} J_0 (\frac{m}{T}) + 
\frac{1}{8}(\frac{\lambda}{16\pi^2}) \left[ ( L +1) m^2 
-T^2 J_1 (\frac{m}{T} ) \right]^2 \,,
\label{F02l}
\ee
where $L\equiv\ln(\mu^2/m^2)$ and we explicitly separated the thermal and non-thermal contributions.
Here and in all related renormalized expressions below, $\mu$ stands for the arbitrary 
renormalization scale introduced by dimensional regularization 
in the $\ms$-scheme, Eq.(\ref{dimreg}), and $\lambda\equiv \lambda(\mu)$. Note carefully that 
${\cal E}_0$ in Eq.(\ref{F02l}) represents a  
{\em finite} vacuum energy term, to be speficied below, that plays a crucial role in our approach 
as will be reexamined in Sec. \ref{pRG}.\\ 
The standard (dimensionless) thermal integrals appearing in Eq.(\ref{F02l}) and below are given by
\be
J_n(x) = 4 \frac{\Gamma[1/2]}{\Gamma[5/2-n]} \: \int_0^\infty dt \frac{t^{4-2n}}{\sqrt{t^2+x^2}}\:
\frac{1}{e^{\sqrt{t^2+x^2}}-1} \,,
\ee
 where  $t=p/T$ and $x=m/T$. Different integrals can be easily related by employing derivatives such as
\be
J_{n+1}(x) = -\frac{1}{2x} \frac{\partial}{\partial x} J_n(x) \,.
\label{Jnrec}
\ee
Also, a high-$T$ expansion\cite{kapusta-gale} such as 
\be
J_0(x) \simeq \frac{16}{45} \pi^4 -4\frac{\pi^2}{3} x^2+ 8\frac{\pi}{3} x^3 +x^4 
\left [\ln \left (\frac{x}{4\pi}\right )+\gamma_E -\frac{3}{4} \right ]  +{\cal O}(x^6)\,,
\label{J0exp}
\ee
is often a rather good approximation as long as $x \lsim 1$, 
i.e., $m\ll T$.\\ 

Next the NNLO contribution involves the additional genuine massive three-loop second integral
in Eq.(\ref{3lbasic}), first calculated in \cite{scal3l}. After algebra, the complete three-loop 
contribution can be expressed as\cite{SPT3l}
\bea
&& F_{3l}= 
-\frac{1}{48}(\frac{\lambda}{16\pi^2})^2 \left[  m^4 \left(5 L^3 +17 L^2 +\frac{41}{2}L -23 
-\frac{23}{12} \pi^2 +2\zeta(3) +c_0 +3(L+1)^2 J_2(\frac{m}{T}) \right) \right. \nnb \\
&& \left. -m^2 T^2 J_1(\frac{m}{T}) \left(12 L^2+28 L-12-\pi^2 -4 c_1 +6(L+1) 
J_2(\frac{m}{T})\right) \right. \nnb \\
&& \left. +T^4 \left( 3 (3 L+4+J_2(\frac{m}{T})) J^2_1(\frac{m}{T})   +
6 K_2(\frac{m}{T}) +4 K_3(\frac{m}{T})\right) \right]
\label{F03l}
\eea
where
\be
c_0 = \frac{275}{12} +\frac{23}{2} \zeta(2)-2 \zeta(3) \simeq 39.429 , \;\;
c_1 = -\frac{59}{8} -\frac{3}{2} \zeta(2) \simeq -9.8424 ,
\ee
and it involves two irreducible, respectively two-loop $K_2(m/T)$ and three-loop $K_3(m/T)$ integrals, 
given explicitly in ref.\cite{scal3l}, reproduced for selfcontainedness in Eqs.(\ref{K2}),(\ref{K3})
in Appendix \ref{K2K3}.

To complete this subsection on basic thermal perturbative expressions, for latter easier reference and comparison purpose 
we also recall the expression of the (massless) PT pressure\cite{Pptphi4,mdeb,Trev} up to NNLO:
\be
\frac{P}{P_0} = 1 -\frac{5}{4}(\frac{\lambda}{16\pi^2}) +5 \frac{\sqrt{6}}{3} (\frac{\lambda}{16\pi^2})^{3/2} 
+\frac{15}{4} (\frac{\lambda}{16\pi^2})^2 \,(\ln \frac{\mu}{2\pi T} +0.40) 
+{\cal O}(\lambda^{5/2}) ,
\label{PPT}
\ee  
where $P_0=(\pi^2/90)T^4$ is the ideal bosonic gas pressure.
\subsection{Perturbatively RG-invariant massive free energy}\label{pRG}
At this stage an important feature is that the (massive) 
free energy is lacking RG invariance. Namely, applying to Eq.(\ref{F02l}), with ${\cal E}_0=0$, the standard RG operator,
\be
\mu\frac{d}{d\,\mu} =
\mu\frac{\partial}{\partial\mu}+\beta(\lambda)\frac{\partial}{\partial \lambda}
+\gamma_m(\lambda)\,m
 \frac{\partial}{\partial m}\;,
 \label{RGop}
\ee 
with $\beta(\lambda)$, $\gamma_m(\lambda)$ given in Eqs.(\ref{beta}),(\ref{gamma}), 
yields a remnant contribution of {\em leading} (one-loop) order: $-(m^4/2) \ln \mu$, for arbitrary $m$. 
Indeed the latter term is not compensated
by lowest orders terms from $\beta(\lambda)$ or 
$\gamma_m(\lambda)$ in Eq.(\ref{RGop}),  
those being at least of next order ${\cal O}(\lambda m^4)$. 
This is a manifestation that perturbative 
RG invariance generally occurs from cancellations between terms from the RG equation 
at order $\lambda^k$ and the explicit $\mu$ dependence at the next order $\lambda^{k+1}$.   
Nevertheless perturbative RG invariance can easily be restored by adding the 
finite vacuum energy term ${\cal E}_0$ to the action, 
although this term is usually ignored, minimally set to zero in the (thermal) 
literature~\cite{SPT3l,htlpt1,HTLPT3loop}.
Following Refs. \cite{prl_phi4,prd_phi4} the easiest way to construct a (perturbatively) RG-invariant
renormalized vacuum energy is to determine ${\cal E}_0$ order by order as a perturbative series 
from the reminder of Eq.~(\ref{RGop}) applied 
on the non RG-invariant finite part of Eq.~(\ref{F02l}):
\be
\mu \frac{d}{d\mu}{\rm {\cal E}_0}(\lambda,m) \equiv -{\rm Remnant}(\lambda,m)= -\mu \frac{d}{d\mu}
[{\cal F}_0({\rm {\cal E}_0\equiv 0})|_{\rm finite}]\; .
\label{rganom}
\ee
Accordingly $ {\cal E}_0$ 
has the form
\be
{\rm {\cal E}_0}(\lambda,m) = -\frac{m^4}{\lambda} \sum_{k\ge 0} s_k  \lambda^{k}\;,
\label{sub}
\ee
where the coefficients $s_k$ can be determined
at successive orders from knowing the 
(single) powers of $\ln \mu$ at order $k+1$ 
(or equivalently the single poles in $1/\epsilon$ of the unrenormalized 
expression)~\cite{rgopt_alphas}.
This procedure leaves non RG-invariant remnant terms 
of higher orders, that may be treated similarly once higher order terms are considered.  
Explicitly, we obtain\cite{prd_phi4}
\bea
& s_0 =\frac{1}{2(b_0-4\gamma_0)} = 8\pi^2\;\nnb \\
&s_1=\frac{(b_1-4\gamma_1)}{8\gamma_0\,(b_0-4\gamma_0)} = -1\; 
\label{s0s1}
\eea
and similarly for the next orders presently relevant,
\bea
& s_2 = \frac{96\pi^2\,(b_0-128\pi^2\left((1+4s_1)\gamma_1-s_0(b_2-4\gamma_2)\right)-41}
{12288\,\pi^4(b_0+4\gamma_0)} = \frac{23+36\zeta[3]}{480\,\pi^2}\simeq 0.01399 \;,\nnb \\
&s_3= \frac{-709 + 12\pi^4 -2628 \zeta(3) -5400 \zeta(5)}{720\,(16\pi^2)^2} .
\label{s2s3}
\eea
The explicit RG coefficients in intermediate expressions emphasizes 
the general form of these results, while the $s_3$ expression is specific to the $N=1$ $\lambda \phi^4$ model. \\
We stress that the previous construction, being only dependent on the renormalization procedure, 
does not depend on temperature-dependent contributions: 
at arbitrary perturbative orders the $s_k$ coefficients are determined 
from the $T=0$ contributions only. Indeed as Eq.(\ref{rganom}) suggests,  
its RHS precisely defines the vacuum energy anomalous dimension, that has been calculated even to five-loop
order for the general $O(N)$ scalar model~\cite{vacanom_kastening}. 
Our independent results for the $s_k$ are fully consistent with \cite{vacanom_kastening}. 
A subtlety is that according to Eq.~(\ref{sub}), $s_k$ is strictly required for
perturbative RG invariance at order $\lambda^k$, but contributes at order $\lambda^{k-1}$.
So at order $\lambda^k$ one may choose minimally to include only $s_0, \cdots s_k$, or  
more completely include also $s_{k+1}$, incorporating in this way higher order RG dependence within the resulting expression.
\section{RGOPT $\lambda \phi^4$ pressure at LO and NLO}
In this section we briefly recall the RGOPT construction, as investigated 
for the $\lambda \phi^4$ model in \cite{prl_phi4,prd_phi4} at LO and NLO, before to extend our approach at the technically more involved NNLO.
We also underline some important features that were not plainly discussed in \cite{prd_phi4}.
After restoring in a first stage perturbative RG 
invariance of the massive free energy, leading to the crucial additional term in Eq.(\ref{sub}), 
one performs on the resulting complete expression the variational modification, according to
\be {\cal F}_0(m^2 \to m^2 \,(1-\delta)^{2a}\,,\;\;\lambda \to \delta \lambda), 
\label{subst1}
\ee
where $m$ is from now an arbitrary variational mass, and the crucial role of the exponent $a$ will be specified just next. 
One then reexpands Eq.(\ref{subst1}) at successive orders, $\delta^k$, at the same order as the original perturbative expression, 
and set $\delta\to 1$ afterwards. This leaves a remnant $m$ dependence at any order $k$, that may be conveniently
fixed by a stationarity prescription~\cite{pms}, 
\be 
\frac{\partial{\cal F}_0^{(k)}}{\partial m}(m,\lambda,\delta=1)|_{m\equiv \t m}  \equiv 0,
\label{OPT}
\ee
thus determining a dressed mass $\t m(\lambda)$ with a ``nonperturbative'' (all order) $\lambda$-dependence.
In OPT~\cite{OPT-LDE} or similarly SPT\cite{SPT} applications, 
the linear $\delta$-expansion has been mostly used, {\it i.e.} assuming $a=1/2$
in Eq.(\ref{subst1}), that corresponds to the ``add and subtract a mass'' intuitive prescription, one mass being
treated as an interaction term. 
Yet it was pointed out that the rather drastic modification implied by Eq.(\ref{subst1}) 
is generally not compatible with RG invariance\cite{prd_phi4}: in contrast 
$a$ can be uniquely fixed~\cite{rgopt_alphas,prd_phi4} by (re)imposing RG invariance now for the variationally
modified perturbative expansion.
Once combined with Eq.(\ref{OPT}), the RG Eq.(\ref{RGop}) takes 
the {\em massless} form
\be
\left[\mu\frac{\partial}{\partial\mu}
+\beta(\lambda)\frac{\partial}{\partial \lambda}\right]{\cal F}_0^{(k)}(m,\lambda,a,\delta=1)=0 ,
\label{RGred}
\ee 
that at leading RG order uniquely fixes\cite{rgopt_alphas,prd_phi4} 
\be
a=\frac{\gamma_0}{b_0} ,
\label{acrit}
\ee
simply in terms of the universal (renormalization scheme independent) first order RG coefficients\footnote{At higher orders one could
generalize the interpolation $(1-\delta)^a $ with $\delta^2$ and higher order terms without spoiling 
the crucial RG properties guaranteed from Eq.(\ref{acrit}). But this would involve extra arbitrary variational parameters 
with no compelling reasons. We thus keep the simpler form Eq.(\ref{acrit}) at successive orders for sensible comparisons.}.

At higher orders,  Eq.(\ref{RGred}) is no longer exactly fulfilled, and can be thus used to determine an
RG-compatible dressed mass, $\t m_{RG}(\lambda,T)$, as a possible alternative to the OPT Eq.(\ref{OPT}). Importantly
Eq.(\ref{acrit}) guarantees in addition that either Eq.(\ref{OPT}) or Eq.(\ref{RGred}) have at least one (essentially unique) 
solution matching the $T=0$ perturbative behavior~\cite{rgopt_alphas,prd_phi4} for $\lambda\to 0$,
{\it i.e.} infrared freedom in the present case: $\lambda(\mu\ll m)\simeq
[b_0 \ln (m/\mu)]^{-1}$. \\
\subsection{LO RGOPT}
For simple illustration let us briefly recall 
the one-loop results\cite{prl_phi4,prd_phi4}, thus considering the LO term in Eq.~(\ref{2lbasic}), including 
solely the first order subtraction term, ${\cal E}_0 = -(m^4/\lambda)s_0$, with $s_0$ given in Eq.(\ref{s0s1}). 
Performing (\ref{subst1}), expanding to leading order $\delta^0$ consistently, 
and taking afterwards $\delta\to 1$ yields
\be
(4\pi)^2 {\cal F}^{\delta^0}_0 = 
-m^4 \left[\frac{1}{2b_0\,\lambda} +\left (\frac{3}{8}+\frac{1}{4}\ln\frac{\mu^2}{m^2} \right) \right] 
-\frac{T^4}{2} J_0 (\frac{m}{T}) .
\label{F0opt0}
\ee
At this leading order Eq.(\ref{RGred}) is satisfied exactly, so that only Eq.(\ref{OPT})
can determine a nontrivial dressed thermal mass. 
It is convenient to introduce first the one-loop renormalized self-energy including 
all the relevant $T$-dependence,  $\Sigma_R$, explicitly\cite{Trev,SPT3l} 
\be
\Sigma_R(m) 
=\gamma_0 \lambda \left[m^2 \left (\ln\frac{m^2}{\mu^2} -1\right )+T^2 J_1\left (\frac{m}{T}\right )\right].
\label{Sigma}
\ee
Then the exact solution of Eq.~(\ref{OPT}), using Eq.(\ref{Jnrec}), is given by the 
self-consistent gap equation: 
\be
\t m^2 = \frac{1}{2}(\frac{b_0}{\gamma_0}) \: \Sigma_R (\t m^2) = (4\pi)^2\,b_0\:\Sigma_R (\t m^2) .
\label{opt0Tex}
\ee
Accordingly $\t m$ is exactly (one-loop) RG-invariant: 
$\lambda\equiv\lambda(\mu)$ being given by the ``exact" (one-loop) running,
\be
\frac{1}{\lambda(\mu)} = \frac{1}{\lambda(\mu_0)} -b_0 \ln \frac{\mu}{\mu_0} ,
\label{g1L}
\ee
the free energy Eq.(\ref{F0opt0}), and therefore $\t m$ in Eq.(\ref{opt0Tex}), only depend on the combination
$1/(b_0\lambda(\mu)) + 1/2 \ln (\mu^2/m^2) $  that is $\mu$-independent.
Let us remark at this stage that some interesting 
qualitative similarities were observed\cite{prd_phi4} between those (one-loop) RGOPT results and the (two-loop) 
2PI resummation approach in \cite{2PI,2PI3loop}, although our construction is basically very different.\\
To get more insight on some properties of the solution of Eq.(\ref{opt0Tex}), one
may conveniently use the high-temperature expansion of the relevant $J_n(x)$, from Eq.(\ref{J0exp})
with $x\equiv m/T$~\footnote{At one-loop order this approximation is valid at
the $0.1\%$ level even for $x \lsim 1$, sufficient for our purpose since the 
RGOPT one-loop solution $\t m/T$ happens to always lies in this range.}.
Accordingly Eq.~(\ref{opt0Tex}) simplifies to a quadratic
equation for $x$, with a unique physical ($x>0$) solution: 
\be
\ds \t x = \frac{\t m^{(1)}}{T} = \pi  \frac{\sqrt{1+\frac{2}{3}\left (\frac{1}{b_0\lambda}+ L_T\right )}-1}
{\frac{1}{b_0\lambda}+ L_T} \simeq \pi\,\sqrt{\frac{2}{3} b_0 \lambda} -\pi\, b_0 \lambda +{\cal O}(\lambda^{3/2})
\label{solopt0T}
\ee
 with $L_T\equiv \ln [\mu\,e^{\gamma_E}/(4\pi T)]$.
Perturbatively at high temperature, the variationally determined mass Eq.(\ref{solopt0T}) 
has the form of
a screening mass, $\t m^2 \sim {\cal O}(\lambda T^2)$, 
but note that this variational mass parameter 
is unrelated to the physical Debye mass~\cite{mdeb} definition. 
The corresponding one-loop RGOPT pressure, from Eq.(\ref{F0opt0}), reads
\be
\frac{P^{(1)}}{P_0}(\t x) = 1 -\frac{15}{4\pi^2}\t x^2 +\frac{15}{2\pi^3}\t x^3
+ \frac{45}{16\pi^4} \left (\frac{1}{b_0\lambda}+L_T\right ) \t x^4 +{\cal O}(\t x^6). 
\label{P1P0}
\ee
Eqs.~(\ref{opt0Tex})-(\ref{P1P0}) clearly involve an all order dependence in $\lambda$,
and RG-invariance is once more manifest 
since from Eq.(\ref{g1L}), $1/(b_0\lambda(\mu)) + L_T $ is $\mu$-independent. 
Accordingly Eqs.~(\ref{solopt0T}) and (\ref{P1P0}) only depend on the single 
parameter $b_0\lambda(\mu_0)$, where $\mu_0$ is some reference scale, typically $\mu_0=2\pi T_0$.

\begin{figure}[h!]
\epsfig{figure=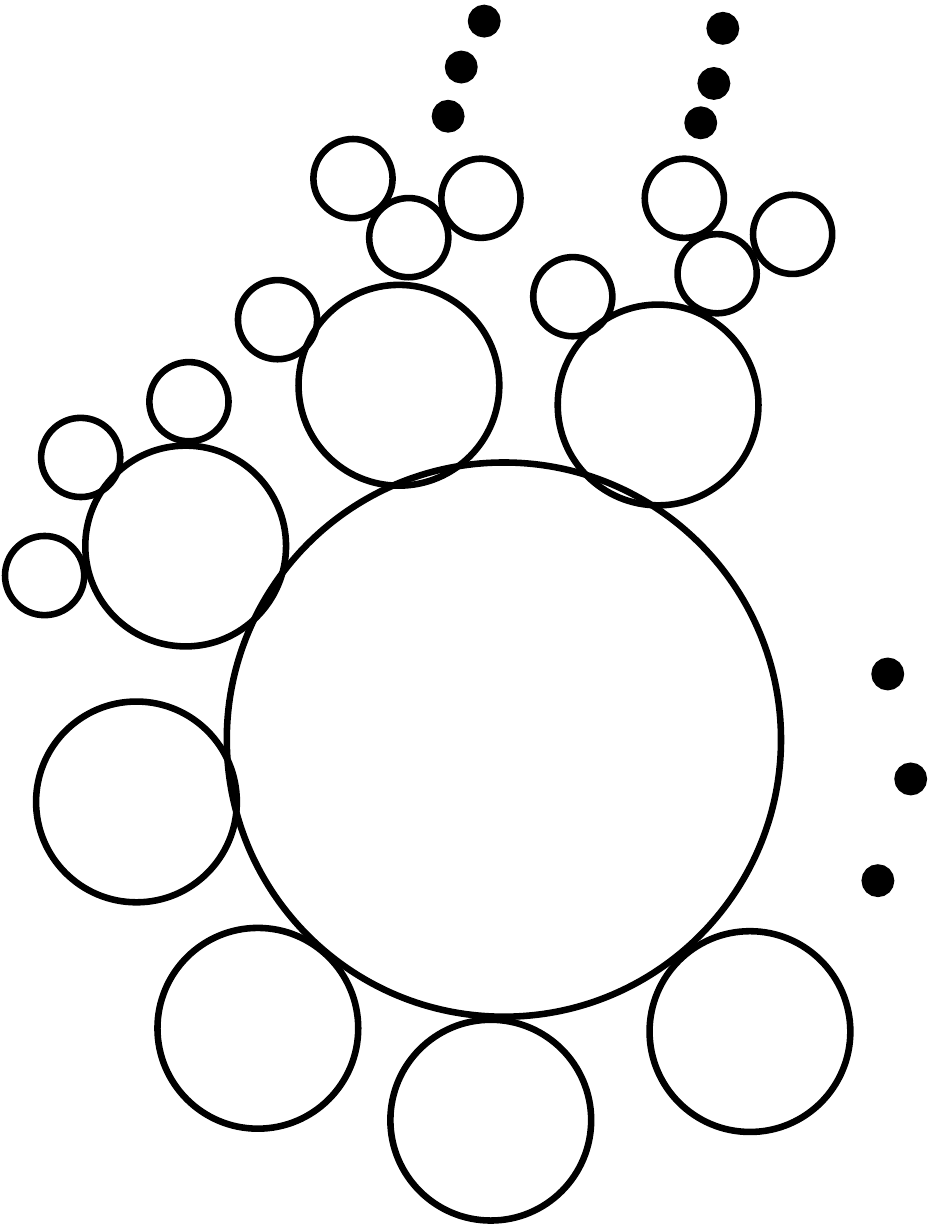,width=3cm,height=3.5cm}
\caption{The graphs being resummed at first nontrival RGOPT order. }
\label{foamLN}
\end{figure}

Expanding perturbatively Eq.~(\ref{P1P0}) one obtains for the first few orders
\be
\frac{P^{(1)}}{P_0} = 1 -\frac{5}{4}\alpha +5 \frac{\sqrt{6}}{3} \alpha^{3/2} +\frac{5}{4}(L_T-6) \alpha^2 
+{\cal O}(\alpha^{5/2})
\label{PP0exp}
\ee
where $\alpha\equiv b_0 \lambda$. Note in particular that Eq.(\ref{PP0exp}) contains 
the nonanalytic term $\sim \lambda^{3/2}$, originating from the
bosonic zero mode resummation, but here readily obtained from RG properties. Expanding at higher orders Eq.(\ref{solopt0T})  
it is easily seen that it entails nonanalytic terms $\lambda^{(2p+1)/2}, p\ge 1$ at all orders. 
Incidentally it is worth mentioning that the higher orders beyond Eq.(\ref{PP0exp}) obtained from 
Eq.(\ref{solopt0T}) correctly reproduce all orders 
of the $O(N$) scalar model large 
$N$-results ({\it e.g.} Eq.~(5.8) of \cite{phi4N}) (once including higher order ${\cal O}(x^6)$ terms, 
not given in Eq.(\ref{J0exp})), 
as can be checked upon identifying the correct large-$N$ $b_0 =1/(16\pi^2)$ value\cite{phi4N}.
Accordingly although the LO RGOPT is essentially built on the very first one-loop graph of Fig.\ref{Fgraph} augmented
by the optimized  RG construction as above described, it happens to correctly resum the whole set of 'foam' graphs as illustrated
in Fig.\ref{foamLN}.\\
\subsection{NLO RGOPT}
The NLO (two-loop) ${\cal O}(\delta^1)$ contribution to the free energy, for $ \delta= 1$, takes 
a rather compact form in terms of $\Sigma_R$ in Eq.~(\ref{Sigma}):
\be
{\cal F}^{\delta^1}_0 =\frac{ {\cal E}^{\delta^1}_0}{(4\pi)^2} + \frac{T}{2}
\sumint_{\bf p} \ln (\omega_n^2+\omega_{\bf p}^2)|_{\mbox{finite}} -\left (\frac{2\gamma_0}{b_0}\right )\frac{m^2}{\lambda}\,\Sigma_R +
\frac{\Sigma_R^2}{2\lambda},
\label{del1compact}
\ee
where the subscript on the integral term means to taking its corresponding finite (renormalized) expression. 
Eq.(\ref{sub}) gives explicitly
\be
{\cal E}^{\delta^1}_0=-m^4 \left( \frac{1}{3 b_0\,\lambda}+\frac{s_1}{3} \right) .
\ee  
The exact two-loop OPT and RG Eqs.~(\ref{OPT}) and (\ref{RGred}) can be written compactly 
as\cite{prd_phi4}~\footnote{Note a typo in Eq.(7.2) of Ref.\cite{prd_phi4}, corrected in Eq.(\ref{OPT2l}), that
does not affect any numerical results.}
\be
f^{NLO}_{\rm OPT}= \frac{2}{3} h \left (-s_1-\frac{1}{b_0\lambda} \right ) +\frac{2}{3} S +\frac{\Sigma^\prime_R}{2m} 
\left (S-\frac{1}{3\lambda}\right )= 0,
\label{OPT2l}
\ee
\be
f^{NLO}_{\rm RG}=  h \left[ \frac{1}{6}+\left (\frac{b_1}{3b_0}-S\right )\lambda \right ] + 
\frac{1}{2} \beta^{(2)}(\lambda) S^2 = 0,
\label{RG2l}
\ee
with $h\equiv (4\pi)^{-2}$, $\beta^{(2)}(\lambda)= b_0\lambda^2 +b_1\lambda^3$,
and we introduced the reduced (dimensionless)
self-energy, 
\be
S(m,\mu,T)\equiv \Sigma_R/(m^2\lambda)
\ee
with, from Eq.~(\ref{Sigma}),
\be
\Sigma^\prime_R\equiv \partial_{m^2}(\Sigma_R)= \lambda (S+m^2 S^\prime) =
\gamma_0\lambda \,[\ln (m^2/\mu^2) -J_2(m/T)] .
\ee
At this stage at NLO in principle we could use three different possible prescriptions to obtain a
thermally dressed mass $\bar m(\lambda,T)$ as function of the coupling,  as investigated in details in \cite{prd_phi4}:  
Either the OPT Eq.(\ref{OPT2l}), or alternatively
the (massless) RG Eq.(\ref{RG2l}), or else 
the full RG, Eq.(\ref{RGop}). The latter is not an independent equation, being a linear
combination of Eq.(\ref{OPT2l}) and (\ref{RG2l}):
\be
f^{NLO}_{\rm full\,RG} \equiv f_{\rm RG}+2\gamma_m(\lambda) f_{\rm OPT} = 0\;.
\label{RGf2l}
\ee
We speculate that if one could calculate to all orders, the solutions of those different prescriptions
would presumably converge towards a unique, nonperturbatively dressed mass $\t m(g,T)$ (as it happens\cite{rgopt1} in the 
large-$N$ limit of the $O(N)$ Gross-Neveu model, where the original perturbative series is known to all orders).
But due to the inherent perturbative truncations those prescriptions give formally different solutions, thus providing
useful variants of the method.
The resulting NLO solutions for $\t m/T$ and $P/P_0$, once reexpanded, are perturbatively consistent 
with Eqs.~(\ref{solopt0T}), (\ref{P1P0})
for the first two order terms, but contain modifications at higher orders\cite{prd_phi4}. 
However, for rather large $g(2\pi T)\lesssim 1$ reference coupling and $\mu=4\pi T$, 
the OPT Eq.(\ref{OPT2l}) no longer gives a real solution: in fact,
the possible occurence of nonreal solutions at higher orders is a recurrent burden of such variational
approach, quite generically expected from the nonlinear $m$ dependence if requiring to solve $\t  m(\lambda,T)$
exactly.
Alternatively using Eq.(\ref{RG2l}) to determine $\t m(\lambda,T)$ gives an unphysical solution at NLO\cite{prd_phi4},
being driven towards the NLO UV fixed point at $\lambda=-b_0/b_1$ 
(an artefact of the scalar model two-loop beta function approximation due to $b_1<0$).
These features lead us to rather consider the third option at NLO, taking the full 
RG Eq.~(\ref{RGop}), explicitly Eq.(\ref{RGf2l}) at NLO, that happens to 
give real solutions at least in a much larger range of coupling 
values~\footnote{In RGOPT applications to thermal QCD\cite{prdCOLD,rgopt_NLOQCD2}, at NLO nonreal solutions 
occur at increasing QCD coupling values for all the prescriptions, an issue that can be circumvented 
at the price of more elaborate prescriptions, based on renormalization scheme changes.}.\\

\begin{figure}[!h]
\epsfig{figure=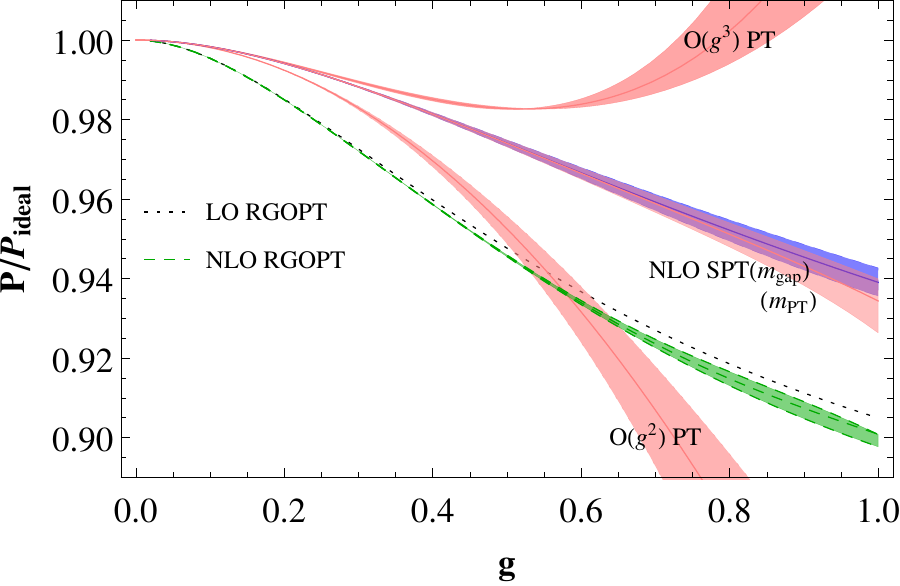,width=8cm,height=6cm}
\caption{LO (dotted lines) and NLO (dashed, green) RGOPT pressure  $P/P_0(g\equiv \sqrt{\lambda/24})$ versus 
${\cal O}(g^2)$ and ${\cal O}(g^{3/2})$ PT, and NLO SPT (for two different prescriptions, see main text). 
The different bands give the scale dependence for $\pi T \le \mu \le4\pi T$. }
\label{Prgopt1and2loop}
\end{figure}
The NLO pressure $P/P_0$ with exact $T$-dependence obtained from 
Eq.~(\ref{RGf2l}), as function of the reference coupling $g\equiv \sqrt{\lambda(2\pi T)/24}$, 
is illustrated in Fig.~\ref{Prgopt1and2loop}, with scale dependence $\pi T \le \mu\le 4\pi T$ using the 
exact two-loop running Eq.(\ref{g2L}). It is compared 
with LO RGOPT, Eq.(\ref{P1P0}), and importantly also with the PT pressure, Eq.(\ref{PPT}), respectively 
truncated at the lowest ${\cal O}(\lambda)\sim g^2$ and next ${\cal O}(\lambda^{3/2})\sim g^3$ orders.
We recall that the latter nonanalytic terms, obtained from resummation of 
a certain class of individually infrared divergent (massless) graphs, are largely responsible for the
poorly convergent, oscillating behavior of successive perturbative orders, as illustrated. In contrast
going from LO to NLO RGOPT appears very stable, despite the fact that both approximations also incorporate 
${\cal O}(\lambda^{3/2})\sim g^3$ contributions (as well as arbitrary higher order contributions as above explained).
Moreover the reduction of the remnant scale dependence as compared to standard PT pressure
is also sizeable, although a moderate residual scale dependence appears at NLO, visible on the figure
for $g \gsim 0.6$.
A more accurate analysis\cite{prd_phi4} shows that at NLO the remnant scale dependence 
reappears first at perturbative order $\lambda^3$.
As above mentioned the exact scale invariance obtained at (one-loop) LO RGOPT (as illustrated in Fig.~\ref{Prgopt1and2loop},
where the dotted line  has no visible width)
results from the peculiar form of the exact running coupling Eq.(\ref{g1L}) perfectly matching Eqs.(\ref{solopt0T}),(\ref{P1P0}), 
since the latter only depends on $b_0$. In contrast at NLO and beyond, $P(\overline m(\lambda),T,\mu)$ 
inevitably has a remnant scale dependence, basically because the subtractions in Eq.(\ref{sub}) only 
guarantee RG invariance up to remnant higher order terms, $\sim m^4 \lambda^2$ at NLO. 
It is a nontrivial consequence of the subsequent variational modification 
in Eq.(\ref{subst1}) that it also preserves this RG invariance at the same level, i.e. up to basically neglected terms
of formally higher orders.

The LO and NLO RGOPT pressures in Fig.~\ref{Prgopt1and2loop}
trivially reproduce the Stefan-Boltzmann limit for $\lambda\to 0$, see Eq.(\ref{P1P0}), 
but one can notice that beyond moderate coupling values they differ 
rather importantly from the PT pressure: this is more pronounced   
in such a plot, conventionally often used\cite{Trev} to illustrate different approximations to
$P(\lambda)$, but where the $\phi^4$ model is not fully specified
by fixing a physical input scale $\mu=T_0$ and corresponding $\lambda(T_0)$ value. 
Note however that upon expressing our results in terms of a more physical mass scale, 
namely solving e.g. Eq.~(\ref{OPT}) for $\t \lambda(m)$ with a resulting $P(m/T)$, 
and inserting the physical Debye screening mass $m(\lambda)$~\cite{mdeb}, correctly reproduces\cite{prd_phi4} 
the first two terms of the standard PT pressure Eq.(\ref{PPT}). Yet the 
RGOPT pressure crucially differs from PT at higher orders, otherwise it would merely reproduce
the same PT behavior and issues. Thus it is important to use the
exact (all order) $\t m(\lambda,T)$ solution from Eq.(\ref{OPT2l}) or (\ref{RG2l}), 
that corresponds to the results shown in Fig.~\ref{Prgopt1and2loop}. (In contrast,
using low orders perturbative reexpansions of the RGOPT $\t m(\lambda)$ solution leads to a behavior 
more similar to the PT pressure, showing large differences between successive 
orders and a larger scale dependence). \\

In Fig. \ref{Prgopt1and2loop} we also compare with the NLO SPT 
results elaborated in Ref.\cite{SPT3l}. Note that all the relevant SPT expressions 
can be obtained consistently by 1) discarding the vacuum energy subtraction ${\cal E}_0$ 
in Eq.(\ref{F02l}); 2) taking $a=1/2$ in Eq.(\ref{subst1}), expanding the result to order $\delta$, and setting $\delta\to 1$; 
and finally 3) 
calculating the variational mass gap, Eq.~(\ref{OPT}), that gives 
explicitly~\footnote{Eq.(\ref{mspt}) is what is called the tadpole prescription in Ref.\cite{SPT3l}.} 
\be
m^2_{\rm SPT} =\Sigma_R ,
\label{mspt}
\ee
see Eq.(\ref{Sigma}), 
to be solved self-consistently for $\t m_{\rm SPT}$. 
Alternatively a simpler prescription was also used in Ref.\cite{SPT3l}, taking instead the 
(NLO) perturbative Debye screening mass\cite{mdeb}:
\be
m^2_{D} = \frac{\lambda}{24}\,T^2 \left(1-\frac{3}{\pi} (\frac{\lambda}{24})^{1/2} \right) , 
\label{msptpt}
\ee
therefore we illustrate these two SPT prescriptions in Fig. \ref{Prgopt1and2loop}. As compared to the two lowest 
orders of standard PT shown, the SPT pressure is significantly more stable and with a 
better remnant scale dependence, that reflects its more elaborate resummation properties.
The SPT pressure values obtained from the two prescriptions are quite close, but using the variational mass gap 
gives a much better remnant scale dependence than using the PT Debye mass. 
The RGOPT remnant scale dependence at NLO is however significantly reduced in comparison:
more precisely for the largest shown (rescaled) coupling $g =1$, the relative $P/P_0$ variation for $\pi T \le\mu \le 4\pi T$
is $\simeq 8\%$, $1.5\%$,  $0.8\%$, and $0.4 \%$ respectively 
for the ${\cal O}(g^3)$ PT, SPT with screening PT mass, SPT with variational mass gap, and RGOPT.
\section{RGOPT $\lambda \phi^4$ pressure at NNLO}
Coming back to the basic free energy in Eq.(\ref{2lbasic}), 
if neglecting the subtraction terms in Eq.(\ref{sub}), the formal lack of RG invariance 
from unmatched $ m^4 \ln\mu$ terms remains relatively screened at 
one- and two-loop orders of thermal perturbative expansions for sufficiently small coupling, since perturbatively 
$ m^4 \sim \lambda^2 $. This can essentially explain why the remnant scale dependence of SPT remains 
quite moderate even at NLO, see Fig. \ref{Prgopt1and2loop}, so that in comparison the NLO RGOPT improvement 
by merely a factor two is not spectacular. But conversely
this can largely explain why a very sizeable scale dependence  
resurfaces for the NNLO SPT pressure\cite{SPT3l}, where the formally same order 
genuine three-loop $\lambda^2$ contributions are considered. 
In contrast the RGOPT scale dependence is expected to further improve at higher orders, at least formally: being built on 
perturbatively restored RG invariance of the free energy at order $m^4 \lambda^k$ for {\em arbitrary} $m$, the 
resulting mass gap exhibits a leading remnant scale dependence as 
$\t m^2 \sim \lambda T^2 (1+\cdots+{\cal O}(\lambda^{k}\ln\mu))$. Thus the dominant 
scale dependence in the free energy, 
coming from the leading term $\sim s_0\, m^4/\lambda$, is expected to appear first only at ${\cal O}(\lambda^{k+1})$.
Nevertheless this expected trend could be largely spoiled, either by large
perturbative coefficients (generically expected to grow at higher orders), or
by the well-known thermal PT issues due to
infrared divergent bosonic zero modes. It is thus important to
investigate more explicitly the outcome of our construction at NNLO, where standard thermal PT starts to 
badly behave, to delineate the RGOPT
scale dependence improvement that can be actually obtained. 
\begin{figure}[h!]
\epsfig{figure=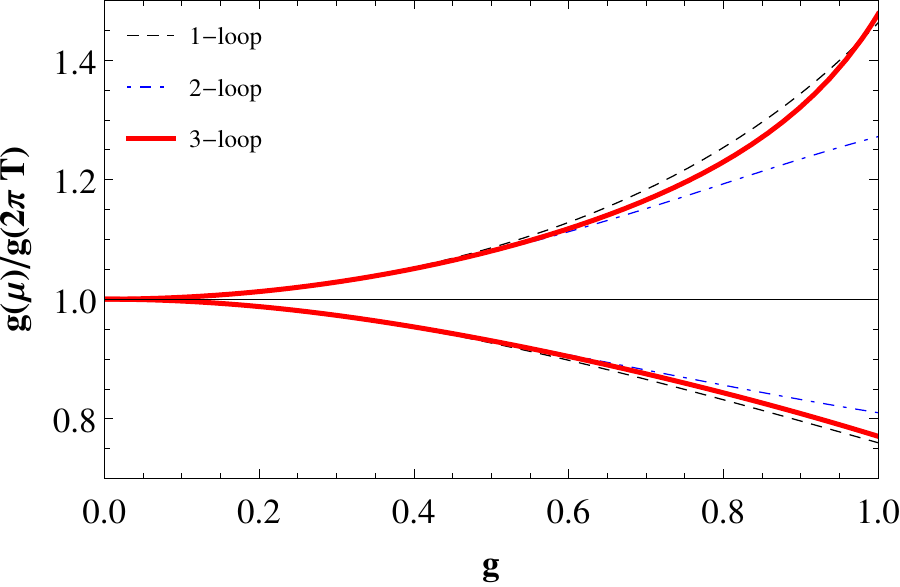,width=7cm,height=5cm}
\caption{Relative scale dependence at successive orders of the running coupling in the $\lambda \phi^4$ model, 
with $g\equiv \sqrt{\lambda/24}$ and $\pi T \le \mu \le 4\pi T $ (from lower to higher values respectively).}
\label{grun}
\end{figure}
Moreover, concerning the $\lambda \phi^4$ model, the peculiar sign alternating beta-function coefficients $b_i$ from
one- to three-loop orders, see Eq.(\ref{bk}), implies 
that considering the running coupling alone, at three-loop order it has a comparatively worse scale dependence than at 
two-loop order, as illustrated in 
Fig.\ref{grun}. This feature tends to partly counteract the benefits of our RG-improved construction, 
when comparing NLO with NNLO\footnote{We expect the corresponding behavior in QCD to be better, 
since the first three QCD beta-function coefficients have the same sign.}.

Applying the variational modification from Eqs.(\ref{subst1}),(\ref{acrit}) to the complete three-loop free energy, sum of 
Eq.(\ref{F02l}) and Eq.(\ref{F03l}), expanded consistently now to order $\delta^2$,
and taking $\delta\to 1$, gives after algebra,
\bea
(4\pi)^2 {\cal F}_0^{\delta^2}  &=& {\cal E}_0^{\delta^2} 
-\frac{m^4}{8}\left( 3+2L + 8 (\frac{\gamma_0}{b_0})^2 
(L+J_2) \right) -\frac{T^4}{2} J_0   \nn \\ 
&& +\frac{\gamma_0}{b_0} \left ( \frac{\gamma_0}{b_0} -\frac{3}{2} \right) \frac{m^2 \Sigma_R}{\gamma_0\,\lambda}  
+\frac{m^2}{32\pi^2 b_0} \Sigma_R \left(L +J_2 \right)  + \frac{\Sigma_R^2}{128\pi^2 \gamma^2_0 \lambda} 
+ F_{3l} ,
\label{del2compact}
\eea
omitting the argument $m/T$ in thermal functions $J_i$, with $L\equiv \ln(\mu^2/m^2)$
and $\Sigma_R$
defined in Eq.(\ref{Sigma}). 
The original ${\cal O}(\lambda^2)$ perturbative contribution, designated by $F_{3l}$ in Eq.(\ref{del2compact}), 
is already defined in Eq.(\ref{F03l}), those terms
being unaffected by Eq.(\ref{subst1}) at $\delta^2$ re-expansion order. 
The subtraction contributions in Eq.(\ref{del2compact}) 
after the modifications from Eq.(\ref{subst1}) read explicitly, using $\gamma_0/b_0=1/6$:
\be
{\cal E}^{\delta^2}_0=-m^4 \left( \frac{14}{81}\frac{s_0}{\lambda} 
+\frac{2}{9} s_1+\frac{1}{3} s_2 
\lambda + s_3 \lambda^2 \right) .
\ee
The explicit expressions at NNLO for the RG and OPT Eqs.(\ref{RGred}), (\ref{OPT}) can be obtained 
straightforwardly from Eq.(\ref{del2compact}) after algebra. They are more involved
than their NLO analogs in Eqs.(\ref{OPT2l}),(\ref{RG2l}) 
and not particularly telling: some relevant expressions are given explicitly in Appendix \ref{solNNLO}. 

Similarly to the NLO, at this stage without examining further constraints 
one may a priori use any of the three possible (not independent) prescriptions to obtain
the NNLO dressed optimized mass $\bar m(\lambda,T)$: namely taking the solution of 
the OPT Eq.(\ref{OPT}), or the massless RG Eq.(\ref{RGred}), or 
the full RG Eq.(\ref{RGop}). 
One may further exploit the freedom to incorporate the highest order subtraction term $s_3$ of Eq.(\ref{s2s3}) or not, 
the latter being formally a three-loop contribution but depending on four-loop RG coefficients, 
thus not necessary for NNLO RG invariance. 
The latter flexibility happens to give a relatively simple handle to circumvent the 
annoyance of possibly nonreal NNLO solutions, that occur only at relatively large couplings 
in the $\lambda\phi^4$ model, quite similarly to what happens at NLO.
The behavior of those solutions for the different prescriptions is detailed in Appendix \ref{solNNLO} 
for completeness. 

The outcome is, in order to maximize the range of coupling and scale $\pi T \le \mu\le 4\pi T$ values 
where real solutions are obtained, it is appropriate to minimally neglect $s_3$ if using the OPT Eq.(\ref{OPT}), and 
incorporating $s_3 \ne 0$ when using the RG Eq.(\ref{RGred}). 
We mention that at NNLO the full RG Eq.(\ref{RGop}) gives no real solutions, 
at least for the relevant scale choice $\mu=4\pi T$ that maximizes the values of $g(\mu)$
for a reference coupling $g\equiv g(2\pi T)$. (Actually one could recover real solutions only if 
truncating Eq.(\ref{RGop}) maximally, keeping only ${\cal O}(g^2)$ terms, but one then loses the crucial perturbative-matching 
properties, so that the corresponding solutions have to be rejected).
Some of these features could be intuitively expected: incorporating higher orders, either via $s_3 \ne0$ 
or taking the more involved exact Eq.(\ref{RGop}), renders the expression even more nonlinear in $m$,  
favoring the occurence of nonreal solutions. But for the massless RG Eq.(\ref{RGred}), provided that it has real solutions, 
incorporating $s_3$ could be expected to give better results, since the subtraction coefficients in Eq.(\ref{s2s3})
entering the pressure are 
originating directly from the RG coefficients. \\

\begin{figure}[h!]
\epsfig{figure=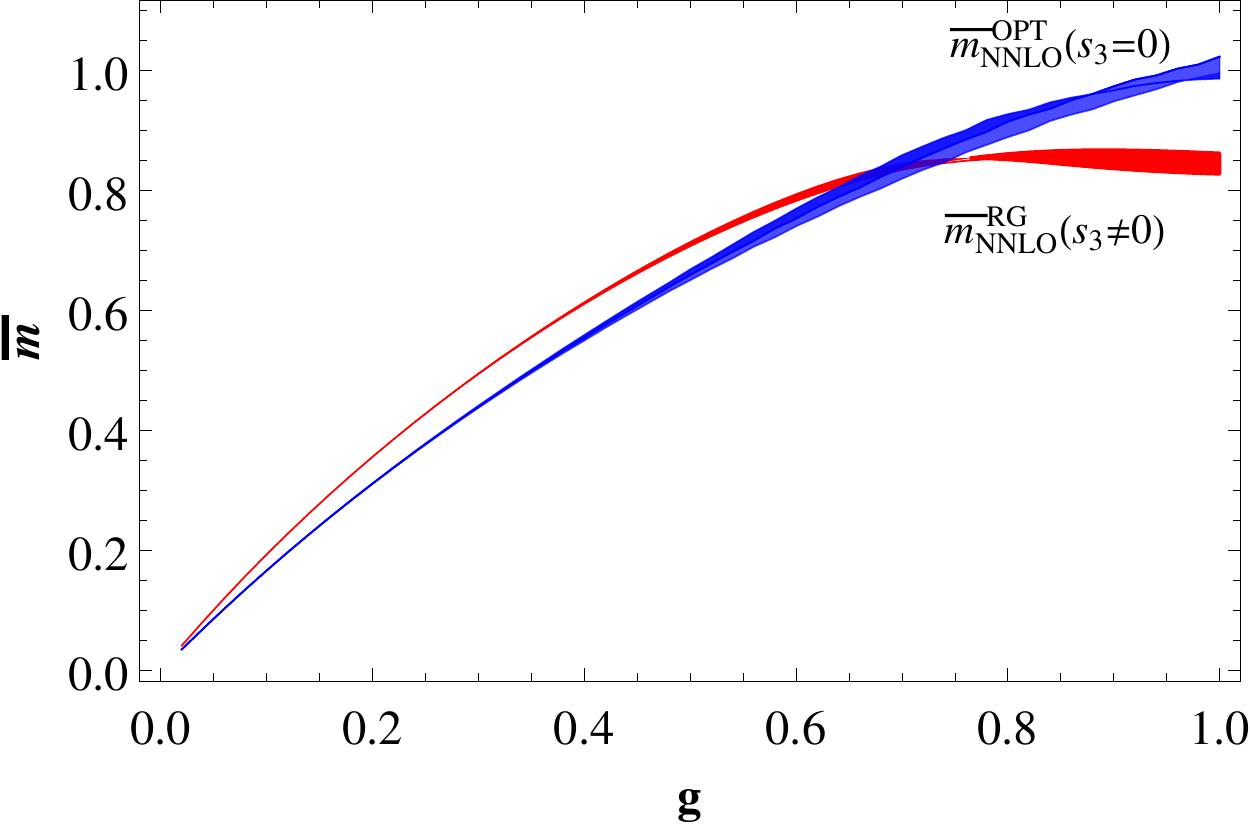,width=8cm}
\caption{NNLO (three-loops) $\overline m$ from 
the OPT Eq.(with $s_3=0$) and RG Eq (with $s_3\ne 0$)
as function of $g\equiv \sqrt{\lambda/24}$,
with scale dependence $\pi T \le \mu \le 4\pi T$.}
\label{mb3l}
\end{figure}

\begin{figure}[h!]
\epsfig{figure=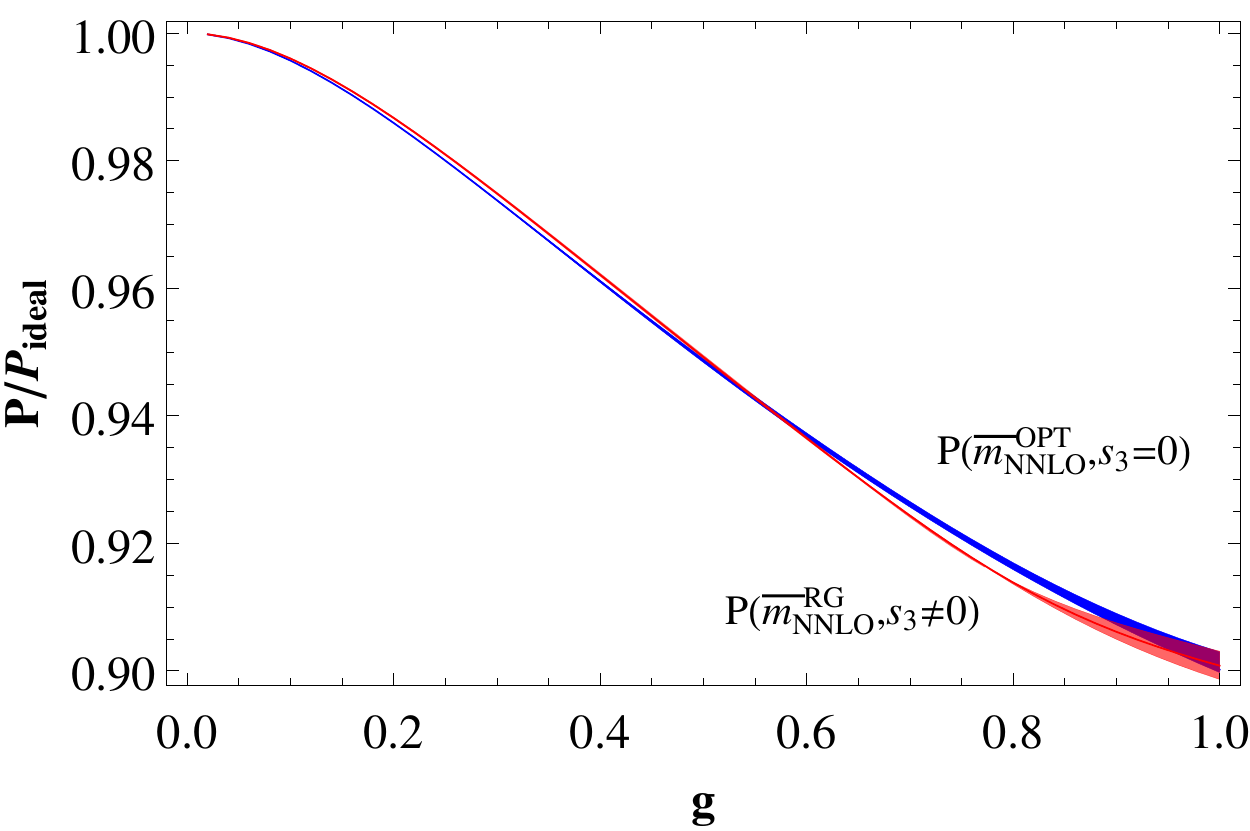,width=9cm}
\caption[long]{Comparison of NNLO pressure $P/P_{ideal}$ for the OPT and RG $\overline m$ prescriptions 
(with $s_3=0$ and $s_3\ne 0$ respectively),
as function of $g\equiv \sqrt{\lambda/24}$
with $\pi T \le \mu \le 4\pi T$. NB: the lower pressure values correspond to the largest $\mu=4\pi T$ 
values due to infrared freedom of $\lambda \phi^4$.}
\label{Poptrg}
\end{figure}

Fig. \ref{mb3l} illustrates as a function of the (rescaled) coupling $g$ our two different prescriptions thus retained at NNLO:
respectively applying to Eq.(\ref{del2compact}) the OPT Eq.(\ref{OPT}), or the RG Eq.(\ref{RGred})
(see also Eq.(\ref{RGNNLO})), with resulting (unique) perturbatively matching exact solutions
$\t m^{OPT}(\lambda,T)$ and $\t m^{RG}(\lambda,T)$ respectively.  As is seen,
despite being rather different functions of the coupling, they
have similar very moderate scale dependence.
Similarly to what happens at NLO the RG solution is generically
giving a slightly better scale dependence than the OPT one, since the former
embeds more directly the perturbative RG properties. 
The previous exact $\t m$ solutions~\footnote{At NNLO we use the exact expressions of all
thermal integrals, in particular Eqs.(\ref{K2}),(\ref{K3}), as their high-$T$ expressions in Eqs.(\ref{K2hT}), (\ref{K3hT})
are not very good approximations.} are then inserted 
within Eq.(\ref{del2compact}) to give the physical pressure.
Fig. \ref{Poptrg} illustrates the corresponding pressures obtained from the two alternative
OPT and RG mass prescriptions. As one can see, despite the quite different $\overline m(g)$
in Fig. \ref{mb3l} the corresponding OPT and RG pressures are very close and similar in shape, and have
comparable very moderate scale dependence. This feature appears as a very good indication of the previously mentioned
expected convergence at higher order of those a priori different prescriptions. \\

\begin{figure}[h!]
\epsfig{figure=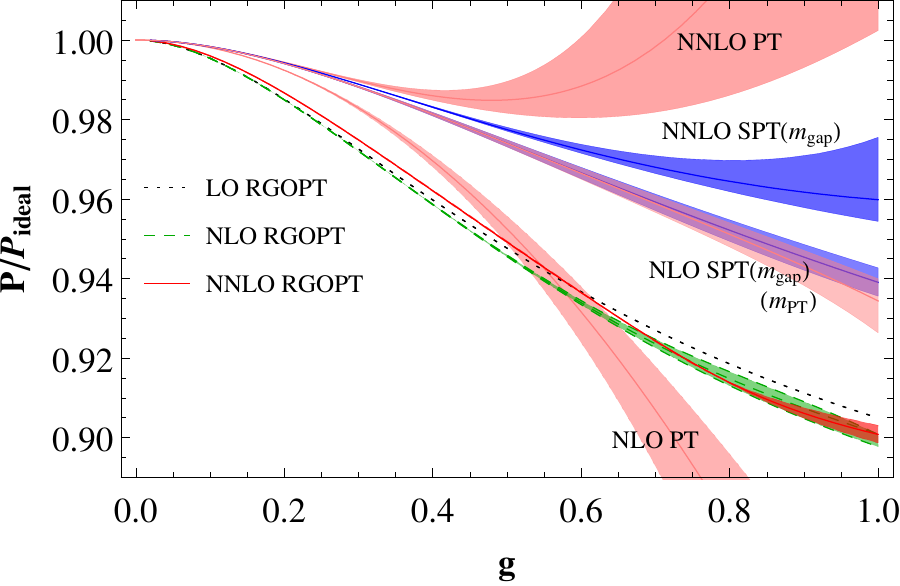,width=10cm}
\caption{RGOPT pressure  at successive LO,NLO,NNLO orders versus NLO,NNLO PT 
and NLO,NNLO SPT pressures, with  $(g\equiv \sqrt{\lambda/24})$, $\pi T \le \mu \le 4\pi T$. 
NB: the lower pressure values correspond to the largest $\mu=4\pi T$ values due to infrared freedom of $\lambda \phi^4$.}
\label{Prgopt3loop}
\end{figure}

Finally we illustrate our main results at successive LO, NLO, and NNLO in Fig. \ref{Prgopt3loop}, 
compared with the PT pressure, Eq.(\ref{PPT}), and SPT pressure at successive orders.
The NNLO PT, with successive terms up to ${\cal O}(\lambda^2)$, has a substantially larger remnant scale dependence
than the ${\cal O}(\lambda^{3/2} )$ PT in Fig.~\ref{Prgopt1and2loop}. Concerning the SPT, at NNLO we only show the results 
from using the mass gap Eq.(\ref{mspt}), following the very same prescription as in \cite{SPT3l}, namely using also Eq.(\ref{mspt}) at
NNLO. (N.B. using the perturbative screeening mass
instead, Eq.(\ref{msptpt}), gives a much larger scale dependence, that we do not illustrate).
In contrast one can see that the RGOPT pressure is remarkably stable from comparing LO to NLO and NNLO, and has a very moderate
remnant scale dependence at NNLO, almost invisible at the figure scale until relatively large $g\simeq 0.8$. 
Actually the improvement from NLO to NNLO becomes only moderate for relatively large coupling values 
$0.9 \lesssim g \lesssim 1$, 
that we understand as the counter-effect from the worse scale dependence of the sole NNLO running coupling as above explained, 
see Fig.~\ref{grun}. Overall the scale dependence is drastically improved as compared to PT and SPT. \\
%
\section{Conclusions and outlook}
We have illustrated in the $\lambda \phi^4$ model up to NNLO how the RGOPT resummation of thermal perturbative expansions, 
consistently maintaining perturbative RG invariance, 
leads to drastically improved convergence and remnant renormalization scale dependence at successive orders, 
as compared with PT and with related thermal perturbative resummation approaches such as SPT. In particular 
RGOPT remains very stable from NLO to NNLO up to relatively large coupling values. We have compared two
different prescriptions a priori available in our framework to defining a resummed dressed mass at a given order, that give
very close and stable results.
As could be intuitively expected, the prescription defining the dressed mass from the RG Eq.(\ref{RGred}), thus more directly
embedding RG properties, appears to give a slightly better remnant scale dependence than the more traditional 
variational mass prescription from Eq.(\ref{OPT}) optimizing the pressure.

The RGOPT has been applied recently at NLO in thermal QCD\cite{rgopt_NLOQCD1,rgopt_NLOQCD2}, yet only in 
the quark sector, thus treating the gluon contributions apart purely perturbatively, due to some
present technical limitations. 
Although such a relatively simple approximation shows a very good agreement with lattice simulations down to 
relatively low temperatures near the pseudo-transition, it is clearly an important next step to extend our approach 
to similarly treat the crucial gluon sector, largely responsible for the poorly convergent weak coupling expansion
of thermal QCD, due to zero mode infrared divergences that start to show up at NNLO. 
Since the very stable NNLO RGOPT properties here demonstrated for the scalar model entails an appropriate 
resummation to all orders of infrared divergent bosonic zero modes, we anticipate similarly
good properties to hold also in the QCD gluon sector, once the technical (computational) difficulties 
to readily adapt the RGOPT approach to the gluon sector will be overcome. 

\newpage
\appendix
\section{Properties of RG and OPT solutions at NNLO}\label{solNNLO}
This appendix examines in some details the properties of the different possible prescriptions at NNLO.
We first give the explicit expression of the NNLO RG Eq.(\ref{RGred}), straightforward to obtain
after algebra from Eq.(\ref{del2compact}),
that takes the rather compact form:
\bea
 f^{NNLO}_{\rm RG} &=& 
-\frac{m^4}{2}\left(1 -6\frac{\gamma_0}{b_0}+ 8 (\frac{\gamma_0}{b_0})^2 \right)
 +\frac{m^2}{16\pi^2 b_0} \left( \Sigma_R -m^2 \gamma_0\lambda(L+J_2) \right) -
\frac{m^2 \Sigma_R}{32\pi^2 \gamma_0}  +G_{3l} \nn \\
&& +\beta^{(3)}(\lambda) \left \{ \frac{m^2 \Sigma_R}{32\pi^2 b_0 \lambda}(L+J_2) +
\frac{\Sigma_R^2}{128\pi^2 \gamma_0^2\,\lambda^2} 
+\frac{2}{\lambda} F_{3l} -m^4 
\left( -\frac{14}{81}\frac{s_0}{\lambda^2} +\frac{s_2}{3} + 2s_3 \lambda \right) \right \} = 0 ,
\label{RGNNLO}
\eea
where 
\bea
\ds
 G_{3l} = 
-\frac{1}{24}(\frac{\lambda}{16\pi^2})^2 && \Big[  m^4 \left(15 L^2 +34 L +\frac{41}{2} 
+6(L+1) J_2 \right) \nnb \\
&&  -m^2 T^2 J_1 \left(24 L+28 +6 J_2 \right) 
 +9T^4 J^2_1 \Big] ,
\label{dmuF03l}
\eea
\be
\beta^{(3)}(\lambda)= \lambda^2 (b_0+b_1 \lambda +b_2 \lambda^2) ,
\ee
with $F_{3l}$ given in Eq.(\ref{F03l}) and $\Sigma_R$ in Eq.(\ref{Sigma}). The OPT Eq.(\ref{OPT}) at NNLO is similarly easily 
obtained from Eq.(\ref{del2compact}), 
but it gives a somewhat more lengthy expression that we thus refrain to give explictly. \\ 
\begin{figure}[h!]
\begin{subfigure}[t]{0.45\textwidth}
        \centering
        \includegraphics[width=\linewidth]{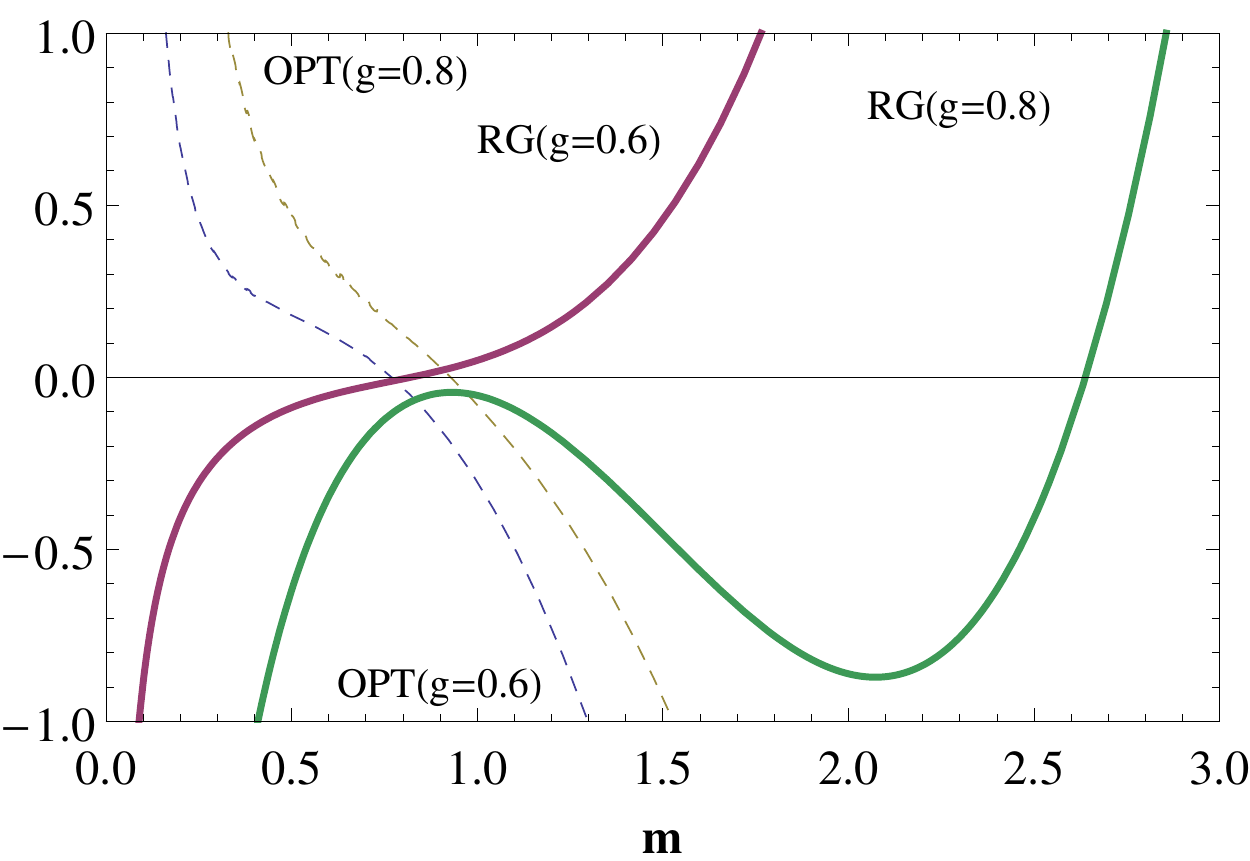} 
    \end{subfigure}

    \begin{subfigure}[t]{0.45\textwidth}
        \centering
        \includegraphics[width=\linewidth]{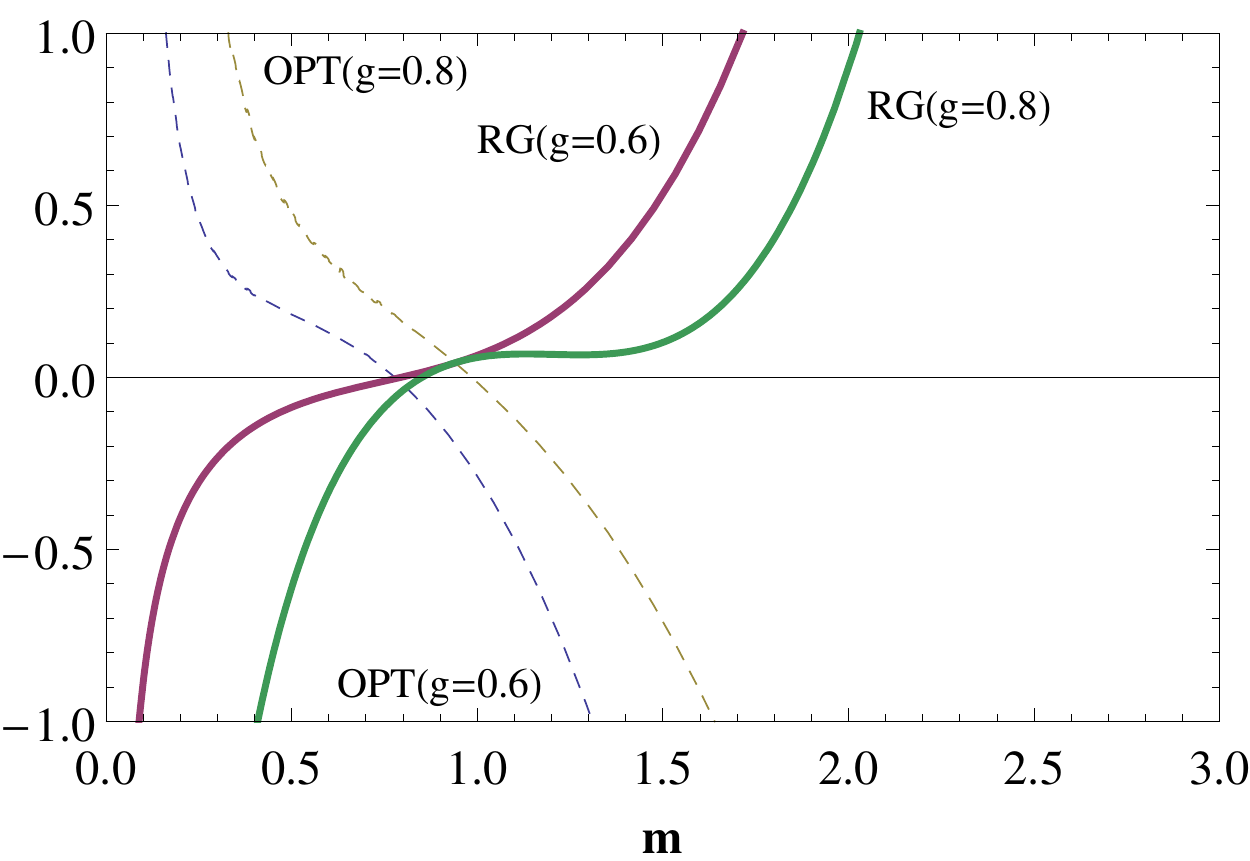} 
    \end{subfigure}
\caption{Top panel: OPT (dashed lines) and RG (thick lines) equations at NNLO
as a function of $m$, for $s_3=0$, 
taking three-loop order running coupling $g(4\pi T)$ ($g\equiv\sqrt{\lambda/24}$) 
for two different values of the reference coupling,
$g(2\pi T)=0.6$ and $g(2\pi T)=0.8$ respectively. Bottom panel: Same captions for 
$s_3 \ne 0$.}
\label{optrgeq}
\end{figure}
Fig.~\ref{optrgeq} illustrates, for $s_3=0$ (top) and $s_3 \ne 0$ (bottom), 
the behavior of these exact RG Eq.(\ref{RGNNLO}) and OPT (\ref{OPT}) respectively,
as function of $m$ for two representative $g\equiv \sqrt{\lambda/24} $ values and for the renormalization scale $\mu=4\pi T$,
giving the largest coupling for a given $g$ input, thus
the most problematic case for sufficiently large $g$. 
For the simpler $s_3 = 0$ prescription, the RG equation 
has a unique perturbative-matching solution 
for $0 < g \lesssim .75$ (for $\mu=4\pi T$), but above this maximal $g$ value 
an inflection point appears that pushes the previous perturbative-matching solution to complex values, 
although being close to real values as can be seen in Fig.~\ref{optrgeq} (bottom).
Note also that another RG solution appears at a higher mass value, but the latter is 
not matching standard perturbative behavior, so it has to be rejected.
Rather similar features are obtained for $\mu=2\pi T$, but the disappearance of the real
perturbative-matching solution is delayed to higher $g\gtrsim 0.95$ values.
In contrast for  $s_3 \ne 0$ a
similar behaviour is obtained but the disappearance of real perturbative-matching
RG solution is delayed to substantially higher $g\gtrsim 1$ values
for any $\mu \le 4\pi T$: accordingly 
for $0 <g \le 1$ and $\pi T \le \mu \le 4\pi T$ the perturbative-matching
solution is real and unique.
The OPT equation has a unique perturbative-matching solution for both 
$s_3 =0$ or $s_3 \ne 0$, and we adopt the simpler minimal choice $s_3=0$ for our NNLO results. 
Notice that despite their very different expressions, the RG and 
OPT equations give $\overline m$ solutions that are rather close to each other. (Even
when the RG solution is not real it is seen to be very close to the OPT real one).\\
Finally note that the full RG Eq.(\ref{RGop}) has no real solutions for $0 < g \le 1$ for $\mu=4\pi T$: 
the latter is not illustrated on Fig.~\ref{optrgeq} but its shape looks quite similar to the RG one
in the range where the latter gives complex solutions (also being close to give a real solution).
\section{RG ingredients and counterterms}\label{appRG}
Our normalization for the beta-function and mass anomalous dimensions are respectively
\be
\beta(\lambda)\equiv \frac{d\lambda}{d\ln\mu}= b_0 \lambda^2 
+b_1 \lambda^3 +b_2 \lambda^4+ \cdots
\label{beta}
\ee
and 
\be
\gamma_m(\lambda) \equiv \frac{d \ln m}{d\ln\mu}= 
\gamma_0 \lambda +\gamma_1 \lambda^2 +\gamma_2 \lambda^3+\cdots
\label{gamma}
\ee
where up to three-loop order~\cite{RGphi4loop} 
\be
(4\pi)^2 b_0 = 3;\;\;  (4\pi)^4 \, b_1 = -17/3 ;\;(4\pi)^6\,  b_2 = \frac{3915 + 2592 \zeta(3)}{216}, 
\label{bk}
\ee
\be
(4\pi)^2 \gamma_0 = 1/2;\;\; (4\pi)^4 \,\gamma_1 = -5/12;\;\; 
(4\pi)^6 \,\gamma_2 = \frac{7}{4} .
\label{gak}
\ee
Next the coupling and mass counterterm are expressed as perturbative 
$\lambda$ and $1/\epsilon$ expansions, with ($d=4-2\epsilon$), 
$\lambda_0 \equiv \mu^{2\eps} \lambda \,Z_\lambda$, $m_0\equiv m Z_m$
(where only the $\lambda^2$ terms are needed at three-loop order):
\be
Z_\lambda = 1 +\frac{b_0}{2\eps}\lambda +
\left[\left (\frac{b_0}{2\eps}\right )^2 +\frac{b_1}{4\epsilon}\right] \lambda^2+
{\cal O}(\lambda^3)
\label{Zlam2l}
\ee
\be
Z_m = 1 +\frac{\gamma_0}{2\epsilon}\lambda +
\left[\frac{\gamma_0(\gamma_0+b_0)}{8\epsilon^2} +\frac{\gamma_1}{4\epsilon}\right] 
\lambda^2+{\cal O}(\lambda^3)
\label{Zm2l}
\ee
The necessary additional vacuum energy counterterms up to three loop order
read\cite{vacanom_kastening} in our normalization conventions:
\be
(4\pi)^2 {\Delta \cal F}_0 = m^4\left[ \frac{1}{4\eps} + (\frac{\lambda}{16\pi^2}) \frac{1}{8\epsilon^2}+ 
(\frac{\lambda}{16\pi^2})^2 \left(\frac{5}{48\epsilon^3} -\frac{5}{72\epsilon^2}+\frac{1}{96\epsilon}\right) \right] . 
\label{Zvac}
\ee
Next, the (exact) two-loop  running coupling used in the numerics is:
\be
\lambda^{\rm (2-loop)}(\mu) = \frac{\lambda(\mu_0)}{ f_W\left (\lambda(\mu_0),\ln \frac{\mu}{\mu_0}\right)}
\label{g2L}\;\;,
\ee
with 
\be
f_W(\lambda,L_\mu) = 1-b_0 L_\mu \lambda + \frac{b_1}{b_0} \lambda \ln \left(f_W \:
\frac{1+\frac{b_1}{b_0} \lambda f^{-1}_W}{1+\frac{b_1}{b_0} \lambda}\right)
=-\frac{b_1}{b_0} \lambda \left\{1 +W\left[-\left (1+\frac{b_0}{b_1 \lambda}\right )\:
e^{-[1+\frac{b_0}{b_1 \lambda} (1-b_0 \lambda L) ]}\right] \right\}\;\;,
\label{fW}
\ee
where $L_\mu\equiv \ln(\mu/\mu_0)$ and $W(x)\equiv \ln (W/x)$ is the Lambert implicit function.
For the range of coupling values illustrated in our main figures, $ g\equiv \sqrt{\lambda/24} \lesssim 1$, 
Eqs.(\ref{g2L}), (\ref{fW}) give not much visible differences with a simpler 
perturbatively truncated  expansion at order $\lambda^3$:
\be
\lambda^{-1}(\mu) \simeq \lambda^{-1}(\mu_0) -b_0 L_\mu  -(b_1 L_\mu) \lambda -\left (\frac{1}{2} b_0 b_1 L_\mu^2\right ) 
\lambda^2 -\left (\frac{1}{2} b_1^2 L_\mu^2 +\frac{1}{3} b_0^2 b_1 L_\mu^3\right ) \lambda^3 +{\cal O}(\lambda^4)\;.
\label{g2Lapprox}
\ee
At three-loop order we used quite similarly an exact integral giving the running coupling $\lambda(\mu)$ 
as a more involved implicit function,
numerically solved for $\mu$ as a function of $\mu_0$. For not too large coupling values there is not much visible
differences with a more common perturbatively truncated running coupling:
\bea
&& \lambda^{\rm (3-loop)}(\mu)= \frac{\lambda(\mu_0)}{f^{3L}}, \nn \\
f^{3L} &=& 1-b_0 L_\mu \lambda -b_1 L_\mu \lambda^2 -(b_0 \frac{b_1}{2} L_\mu^2 +b_2 L_\mu) \lambda^3 \nn \\ +
&& \frac{L_\mu}{6b_0^2} \left( 6 b_1 (b_1^2  - 2 b_0 b_2) - 3 b_0^2  (b_1^2  + 2 b_0 b_2) L_\mu - 2 b_0^4  b_1 L_\mu^2 \right) 
 \lambda^4 .
\label{g3L}
\eea

\section{Two- and three-loop irreducible integrals}\label{K2K3}
For completeness we reproduce here the explicit expressions of the thermal three-loop massive integrals $K_2$ and $K_3$, 
originally calculated in \cite{scal3l}, and 
entering Eq.(\ref{F03l}):
\be
   K_2(\frac{m}{T})  = -\frac{32}{T^4}\int_{0}^{\infty}dp\, p \ \frac{ n(E_p)}{E_p}\int_{0}^{p}dq \, q \ 
   \frac{ n(E_q)}{E_q}\int_{p-q}^{p+q}dk\, k \sum_{\sigma=-1,+1}f_2(E_\sigma,k) ,
\label{K2}
\ee
where
\bea
  f_2(E,k) &= \left(\frac{E^2-M_k^2}{E^2-k^2}\right)^{\frac{1}{2}}\ln \frac{\left(E^2-k^2\right)^{\frac{1}{2}}+\left(E^2-M_k^2\right)^{\frac{1}{2}}}
  {\left(E^2-k^2\right)^{\frac{1}{2}}-\left(E^2-M_k^2\right)^{\frac{1}{2}}} \ \ \ \ , \ \  k^2<E^2-4m^2 \nn \\ 
    &= 2\left(\frac{M_k^2-E^2}{E^2-k^2}\right)^{\frac{1}{2}}\arctan \left(\frac{E^2-k^2}{M_k^2-E^2}\right)^{\frac{1}{2}} \ \ \ \ , \ \  E^2-4m^2<k^2<E^2 \nn \\
    &= \left(\frac{M_k^2-E^2}{k^2-E^2}\right)^{\frac{1}{2}}\ln \frac{\left(M_k^2-E^2\right)^{\frac{1}{2}}
    +\left(k^2-E^2\right)^{\frac{1}{2}}}{\left(M_k^2-E^2\right)^{\frac{1}{2}}-\left(k^2-E^2\right)^{\frac{1}{2}}} \ \ \ \ , \ \ E^2<k^2 
\eea
and 
\bea
  M_k^2 &=& 4m^2+k^2 , \;\; k\equiv |{\bf p} +{\bf q}| ,\nn \\
    E_\sigma(p,q)& =&\sqrt{p^2+m^2}+\sigma \sqrt{q^2+m^2} .
\eea
\bea
&&  K_3(\frac{m}{T})  = \frac{96}{T^4}\int_{0}^{\infty}dp\, p \ \frac{n(E_p)}{E_p}\int_{0}^{p}dq\,q \ 
\frac{ n(E_q)}{E_q} \int_{0}^{q}dr\, r \ \frac{ n(E_r)}{E_r} \nn \\
&& \times \sum_{\sigma,\tau=-1,+1} \{ f_3(E_{\sigma\tau},p+q+r)-f_3(E_{\sigma\tau},p+q-r)-
f_3(E_{\sigma\tau},p-q+r)+f_3(E_{\sigma\tau},p-q-r) \} ,
\label{K3}
\eea
where
\bea
    f_3(E,p) &= p \ln \frac{m^2-E^2+p^2}{m^2}+2(m^2-E^2)^{\frac{1}{2}}\arctan \frac{p}{(m^2-E^2)^{\frac{1}{2}}} \ \ \ , \ E^2<m^2 \nn \\
    &= p \ln \frac{\big| E^2-m^2-p^2 \big|}{m^2}+\left(E^2-m^2\right)^{\frac{1}{2}} 
    \ln \frac{(E^2-m^2)^{\frac{1}{2}}+p}{\big|(E^2-m^2)^{\frac{1}{2}}-p\big|} \ \ \ , \ E^2>m^2
\eea
and 
\be
    E_{\sigma\tau}(p,q,r) = \sqrt{p^2+m^2}+\sigma \sqrt{q^2+m^2} +\tau \sqrt{r^2+m^2} \ .
\ee
In the limit $x\equiv m/T \rightarrow 0$ Eqs.(\ref{K2}), (\ref{K3}) can be expressed analytically as:
\be
    K_2(x) \simeq  \frac{(4\pi)^4}{72}\left(\ln x +\frac{1}{2} +\frac{\zeta'(-1)}{\zeta(-1)} \right)-372.65\,x \left(\ln x +1.4658\right) , 
    \label{K2hT}
    \ee
    \be
    K_3(x) \simeq   \frac{(4\pi)^4}{48}\left( -\frac{7}{15} +\frac{\zeta'(-1)}{\zeta(-1)} -\frac{\zeta'(-3)}{\zeta(-3)} \right) +1600.0\, x \left(\ln x +1.3045\right) .
     \label{K3hT}
\ee
For the numerical results we rather use the exact expressions Eqs.(\ref{K2}), (\ref{K3}).

\end{document}